\documentclass[
 reprint,
 amsmath,amssymb,
 aps,
 prl,
]{revtex4-1}

\usepackage{graphicx}% Include figure files
\usepackage{dcolumn}% Align table columns on decimal point
\usepackage{bm}% bold math
\usepackage{subfig}
\usepackage{xcolor}
\usepackage{nicefrac}
\usepackage{chemfig}
\newcommand{\hl}[1]{#1}
\begin{document}

%\preprint{APS/123-QED}

\title{Data Mining for \hl{Three-Dimensional} Organic Dirac Materials: \\Focus on Space Group 19}
\author{R. Matthias Geilhufe$^1$}
\email{Geilhufe@kth.se}
\author{Stanislav S. Borysov$^1$}
\author{Adrien Bouhon$^2$}
\author{Alexander V. Balatsky$^{1,3,4}$}
\affiliation{$^1$Nordita, Center for Quantum Materials, KTH Royal Institute of Technology and Stockholm University, Roslagstullsbacken 23, SE-106 91 Stockholm, Sweden\\
$^2$Department of Physics and Astronomy, Uppsala University, Box 516, SE-751 20 Uppsala, Sweden\\
$^3$Institute for Materials Science, Los Alamos National Laboratory, Los Alamos, NM 87545, USA\\
$^4$\hl{ETH Institute for Theoretical Studies, ETH Zurich, 8092 Zurich, Switzerland}}

\date{\today}

\begin{abstract} 
We combined the  group theory and data mining approach within the Organic Materials Database that leads to the prediction of stable Dirac-point nodes within the electronic band structure of three-dimensional organic crystals. We find a particular space group $P2_12_12_1$ ($\#19$) that is conducive to the Dirac nodes formation. We prove that nodes are a consequence of the orthorhombic crystal structure. Within the electronic band structure, two different kinds of nodes can be distinguished: 8-fold degenerate Dirac nodes protected by the crystalline symmetry and 4-fold degenerate Dirac nodes protected by band topology. Mining the Organic Materials Database, we present band structure calculations and symmetry analysis for 6 previously synthesized organic materials. In all these materials, the Dirac nodes are well separated within the energy and located near the Fermi surface, which opens up a possibility for their direct experimental observation. 
 \end{abstract}
 
\keywords{Dirac materials, organometallics, electronic structure, data mining}
\maketitle

% why Dirac materials
Recently, we have witnessed growing interest in the research community in the Dirac materials where the low-energy excitations behave as massless Dirac fermions ~\cite{wehling2014}. Among the most prominent examples are the two-dimensional material graphene~\cite{abergel2010properties}, the surface of bulk topological insulators~\cite{fu2007topological} like Pb$ _x$Sn$_{1-x}$Te~\cite{tanaka2012experimental,geilhufe2015effect}, Dirac-line materials~\cite{geilhufe20163,Kim2015,Yamakage2016} and Weyl semimetals like TaAs~\cite{lv2015experimental}. 
%They are useful and interesting. This materials can be viewed as ``high energy lab on the tabletop''~\cite{}. Dirac materials offer potential applications in \dots [general motivation blah]
% why organic Dirac materials
To date, the strong focus within electronic Dirac materials lies in the inorganic crystals.  The class of organic crystals remains rather unexplored and only a few organic Dirac materials are known. One prominent example is the quasi two-dimensional charge transfer salt $\alpha$-(BEDT-TTF)$_2$I$_3$ which shows a tilted Dirac cone located at the Fermi energy under high pressure~\cite{Katayama2006}. At the same time, organic crystals offer a high potential for technological applications due to low production costs, elastic properties (flexible electronics) and the opportunity to build large-area devices~\cite{lussem2013doping}. We therefore will focus on exploring the space of organic materials with the specific goal of identifying Dirac materials. 

% search strategy: Data mining
Since the crystal structure plays a crucial role for hosting Dirac nodes, attempts of identifying organic Dirac materials so far are mainly based on variations of already known Dirac materials. In two dimensions, this can be done by starting with the graphene structure and replacing the carbon atoms by more complex organic molecules~\cite{wang2013prediction}. A similar strategy was also discussed with respect to $\alpha$-(BEDT-TTF)$_2$I$_3$~\cite{choji2011zero,morinari2014possible}. 
To go beyond this approach, we adopt a new strategy for the search for Dirac nodes in the class of three-dimensional organic crystals. We performed a data mining study on the basis of 5217 electronic Kohn-Sham band structures calculated using Density Functional Theory (DFT)~\cite{Hohenberg1964,Kohn1965} and stored within the Organic Materials Database (OMDB)~\cite{borysov2016}. We looked for {\it isolated} linear crossings, i.e., where no other bands can be found besides the crossing withing the corresponding energy range. Although all considered organic materials were previously synthesized, little attention has been paid to the electronic structure for most of them so far. %Data mining represents a modern approach within materials science~\cite{sarmiento2015prediction,klintenberg2014computational} sometimes referred to as materials informatics~\cite{rodgers2006materials,ferris2007materials}. 

% search strategy: Group theory
To achieve stable Dirac points within the electronic structure, symmetry or topological protection needs to be present. In this connection, crystals with non-symmorphic space groups have been widely discussed~\cite{Young2015, schoop2015dirac,yang2016} where the key role is played by high-dimensional irreducible representations at the Brillouin zone boundary. As a consequence of the Wigner-Eckart theorem, the degeneracy of an electronic state is equal to the dimension of an irreducible representation. In the context of crystals, these degeneracies were discussed in great detail during the 1960s~\cite{zak1960method}. Recently, with a reinterpretation, degenerate electronic states in crystals again attracted attention as a host to  unconventional fermions as low-energy excitation~\cite{bradlyn2016beyond}. 
There is also a second opportunity of hosting Dirac crossing as accidental crossings~\cite{herring1937}, which are protected by band topology. Such crossings can be found for instance in crystals with the monoclinic space group $P2_1/c$ ($\#14$)~\cite{geilhufe20163,wieder2016spin}. In crystals with this space group, electronic energy bands are sticking together in groups of four bands. As reported in Ref.~\cite{geilhufe20163}, for each of these groups, three topologically different orderings of electronic states can be found at the $\Gamma$-point within the Brillouin zone---a trivial phase and two different line-node phases. Within the space group $P2_12_12_1$ ($\#19$), an almost similar situation is present, with the difference that at least one crossing has to occur along one of the paths $\overline{\Gamma X}$, $\overline{\Gamma Y}$ or $\overline{\Gamma Z}$ within the Brillouin zone. This group will be discussed below in Results.

% what we have done
In this paper, we report results of  a combined study using abstract group theory and data mining within the Organic Materials Database (OMDB)~\cite{borysov2016}. We point out the first real material examples in the class of three-dimensional organic crystals hosting isolated Dirac nodes in the electronic structure: C$_6$H$_7$ClO$_3$~\cite{doi:10.1021/ol9016657}, C$_{10}$H$_{10}$Br$_{2}$Cl$_{3}$NO$_{2}$~\cite{bae2010organocatalytic}, C$_{12}$H$_{13}$NO$_{2}$~\cite{dong2005asymmetric}, C$_{13}$H$_{12}$N$_{2}$O~\cite{butin2010furan}, C$_{9}$H$_{10}$F$_{3}$NO~\cite{vsterk2006highly}, %C$_{11}H$_{13}NO$_{2}S~\cite{evans2007double}, 
and C$_{10}$H$_{12}$BrNO ~\cite{eloi2010cationic}. It will be shown that the found Dirac nodes are a consequence of the orthorhombic crystal structure of the space group $P2_12_12_1$ (\#19). Within the band structure of the materials, two different kinds of nodes can be distinguished: 8-fold degenerate Dirac nodes protected by the crystalline symmetry and 4-fold degenerate tilted Dirac nodes protected by the band topology.
%--------------------------------
\section*{Results}
%--------------------------------

%--------------------------------
\subsection*{Data mining and electronic structure calculations}
%--------------------------------
As reported in Ref.~\cite{borysov2016}, most three-dimensional organic crystals are insulating. However, the doping of organic materials is extensively studied, opening the opportunity of shifting the Fermi level into the valence band ($p$-doping) or conduction band ($n$-doping)~\cite{lussem2013doping,Pfeiffer200389}. \hl{Even though it is chemically more difficult to achieve in comparison to inorganics, for example, due to purification effects of organics, it was successfully implemented within OLEDs }\cite{Zhou2001,Blochwitz1998}\hl{, solar cells} \cite{hoppe2004organic,drechsel2005efficient} \hl{or thermoelectric materials} \cite{kim2013engineered}.  Therefore, we searched for isolated linear crossings in a neighborhood of \hl{0.5~eV} above the lowest unoccupied electronic state and below the highest occupied electronic state within the Kohn-Sham band structures stored in the OMDB by explicitly focusing our attention to materials with the space group $P2_12_12_1$ ($\#19$). For the 6 most promising structures tabulated in Table~\ref{tab:materials}, additional refined DFT calculations were performed (see Methods for more details). 

\begin{figure*}[t!]
\raggedright
%\centering
\subfloat[Orthorhombic unit cell and single molecule of C$_{6}$H$_{7}$ClO$_{3}$. Grey denotes carbon, red oxygen, blue hydrogen and green chlorine\label{f1b}]
 {\includegraphics[height=4.7cm]{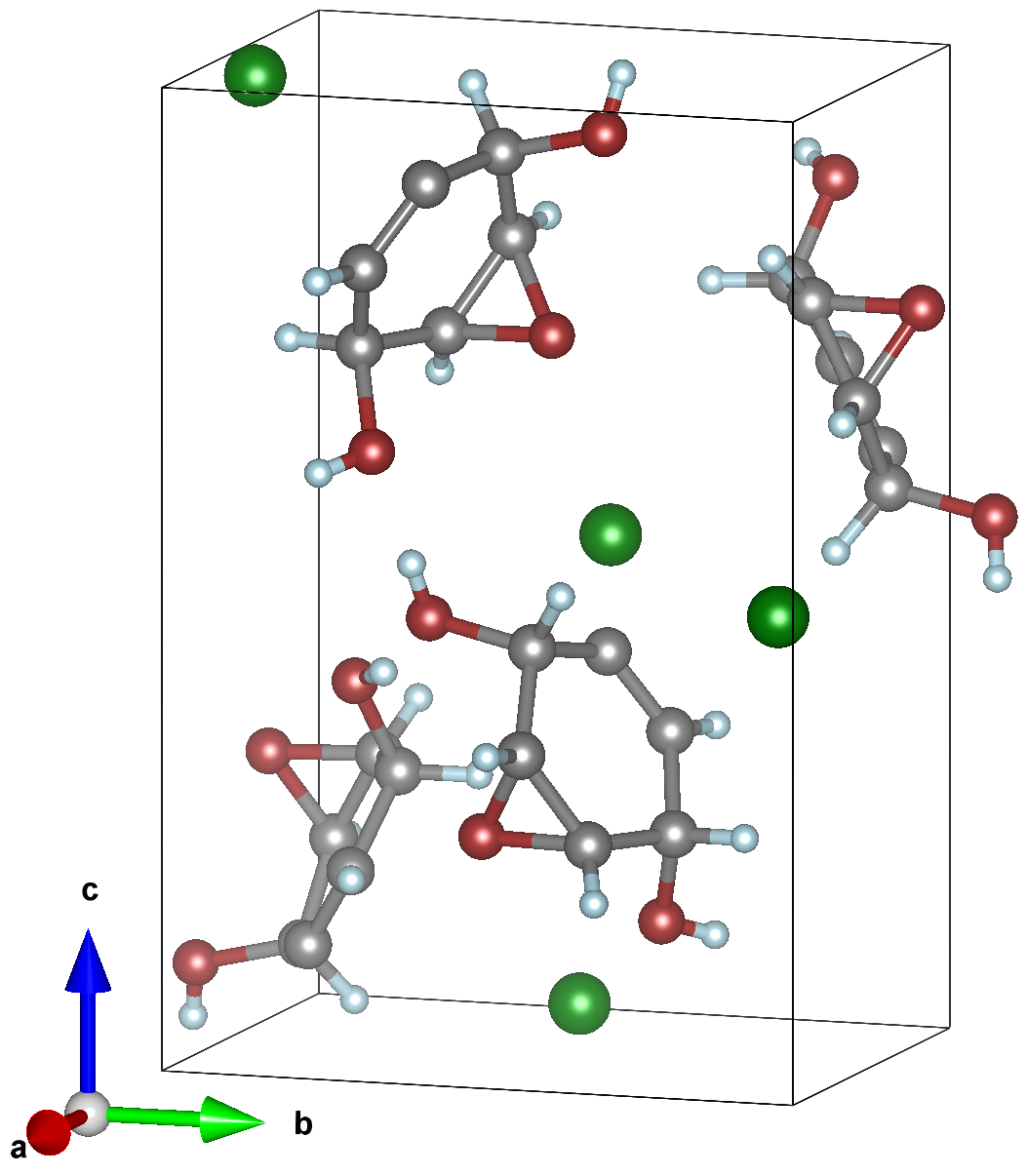}
 \definecolor{Cl}{HTML}{008000}
\definecolor{O}{HTML}{A52A2A}
\hspace{0.2cm}
%\raisebox{2cm}{\chemfig[][scale=0.75]{*6((-\textcolor{Cl}{Cl})=([,0.6]-H)-([8,0.6]-\textcolor{O}{O}-[7,0.6]H)([6,0.6]-H)-([:95,0.8]-\textcolor{O}{O})([8,0.6]-H)-([:20,0.6]-)([2,0.6]-H)-([3,0.6]-\textcolor{O}{O}-[4,0.6]H)([5,0.6]-H)-)}}
\raisebox{1.4cm}{\includegraphics[height=2.2cm]{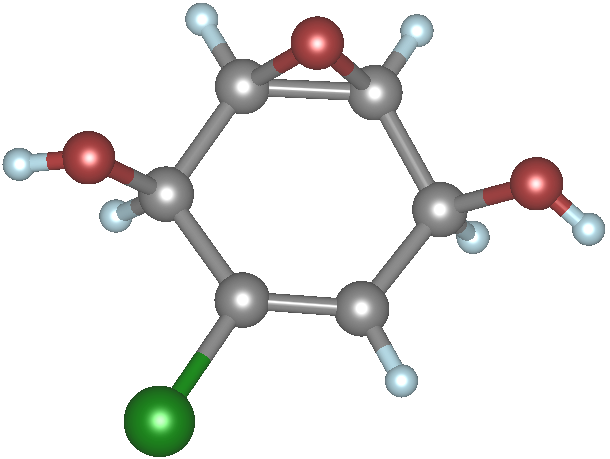}}
}
\hspace{0.2cm}
\subfloat[Brillouin zone path for band structure calculation in Fig.~\ref{f1d}\label{f1a}]
 {\includegraphics[height=4.5cm]{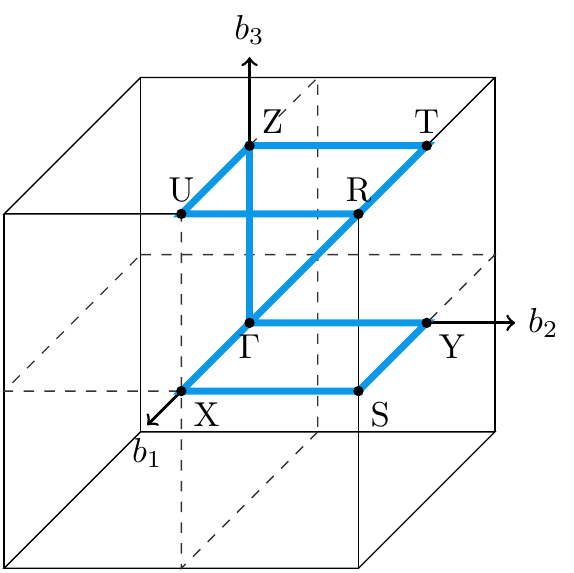}}\hspace{0.2cm}
 \subfloat[Energy dispersion in $\vec{b}_2$-$\vec{b}_3$-plane around the $R$ point\label{f1c}]{\includegraphics[height=3.8cm]{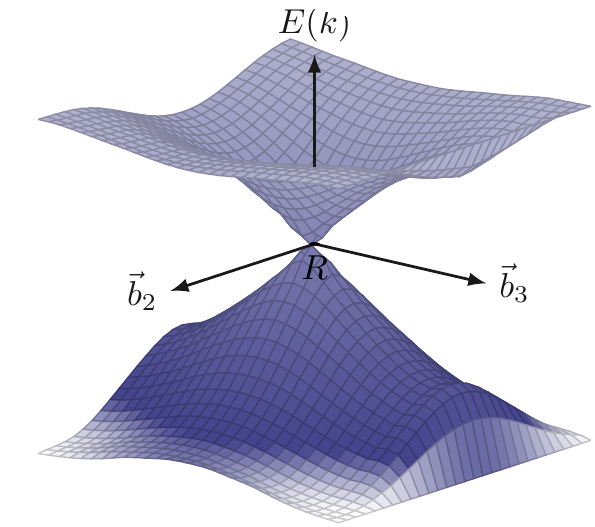}}
 \\
 \subfloat[Band structure and density of states. Crossings highlighted in green circles are 8-fold degenerate Dirac-points protected by crystalline symmetry and crossings in red circles are 4-fold degenerate Dirac points protected by band topology\label{f1d}]{\includegraphics[height=5.5cm]{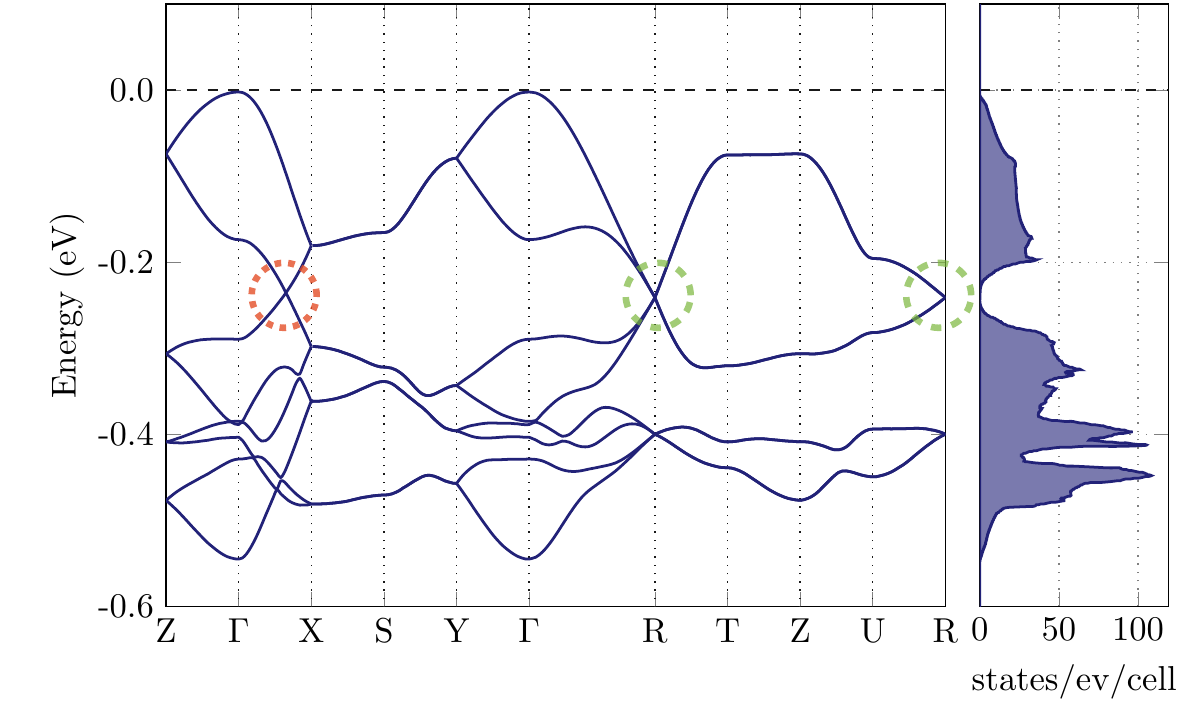}}\hspace{0.5cm}
 \subfloat[Energy dispersion along $\overline{\Gamma X}$, $\overline{\Gamma Y}$, $\overline{\Gamma Z}$ and irreducible representations\label{f1e}]{\raisebox{.4cm}{\includegraphics[height=5.1cm]{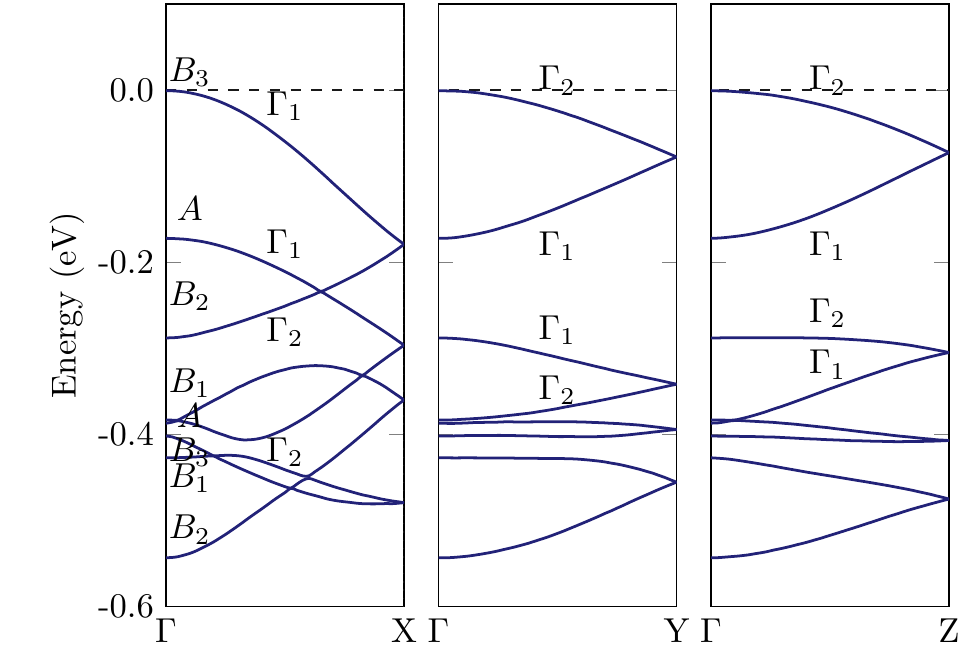}}}\\
\caption{{\bf Crystal structure and electronic structure of the orthorhombic organic crystal C$_{6}$H$_{7}$ClO$_{3}$.}\label{fig:material_0}}
\end{figure*}

As a representative, we further concentrate on the discussion for the material C$_{6}$H$_{7}$ClO$_{3}$ (the electronic and molecular structures for the other 5 materials can be found in the supplementary material). Its orthorhombic crystallographic unit cell is build up of four copies of C$_{6}$H$_{7}$ClO$_{3}$ molecules as shown in Fig.~\ref{f1b}. The calculated band structure along several high symmetry paths within the Brillouin zone (Fig.~\ref{f1a}) is plotted in Fig.~\ref{f1d}. The isolated linear crossings occur at the high-symmetry point $R$ (green dashed circles in Fig.~\ref{f1d}) as well as along the high-symmetry path $\overline{\Gamma X}$ (red dotted circles in Fig.~\ref{f1d}). A picture of the linear energy dispersion within the $\vec{b}_2$-$\vec{b_3}$-plane within reciprocal space in the vicinity of the $R$-point can be seen in Fig.~\ref{f1c}. The linear crossings are well separated within the energy and as a consequence the electronic density of states grows quadratically with the energy in the vicinity of the three-dimensional Dirac crossing ($n(E)\sim E^{d-1}$~\cite{wehling2014}, where $d=3$) as can be seen in Fig.~\ref{f1d}. 

However, to justify the claim of the found materials being Dirac materials, a protection of the crossings by crystalline symmetry is necessary. Indeed, we find that the nature of the crossings forming Dirac nodes in the spectra can be explained within the framework of group theory below. 

\begin{table}[!h]
\centering
\caption{{\bf Data-mined organic Dirac materials with the space group $19$.} The first four materials show crossings in the valence band, the latter two have crossings in the conduction band. References to the papers where the synthesis of the materials was reported are given in the last column.  \label{tab:materials}}
\begin{tabular}{rllll} 
\hline
$\#$ & OMDB ID & COD ID & sum formula & ref.\\
%\midrule
\hline
1 & \textbf{3407} & \textbf{1504134} & C$_{6}$H$_{7}$ClO$_{3}$ &~\cite{doi:10.1021/ol9016657} \\
2 & \textbf{4071} & \textbf{1503207} & C$_{10}$H$_{10}$Br$_{2}$Cl$_{3}$NO$_{2}$ &~\cite{bae2010organocatalytic} \\
3 & \textbf{3180} & \textbf{1506714} & C$_{12}$H$_{13}$NO$_{2}$ &~\cite{dong2005asymmetric} \\
4 & \textbf{5010} & \textbf{7153350} & C$_{13}$H$_{12}$N$_{2}$O &~\cite{butin2010furan} \\
\hline 
5 & \textbf{3617} & \textbf{1506293} & C$_{9}$H$_{10}$F$_{3}$NO & ~\cite{vsterk2006highly}\\ 
%6 & \textbf{2861} & \textbf{4022697} & C$_{11}$H$_{13}$NO$_{2}$S & ~\cite{evans2007double} \\
6 & \textbf{690} & \textbf{4066669} & C$_{10}$H$_{12}$BrNO & ~\cite{eloi2010cationic} \\ 
\hline
\end{tabular} 
\end{table}

%--------------------------------
\subsection*{Group theory analysis}
%--------------------------------

The space group $P2_12_12_1$ ($\#19$) itself (hereinafter denoted by $\mathcal{G}$) is an infinite group having the group of pure translations $\mathcal{T}$ as an infinite, normal and Abelian subgroup. The point group of the lattice $\mathcal{G}_0$, i.e., the group of all rotational parts of the space group elements is given by $222$ ($D_{2}$). The factor group $\mathcal{G}/\mathcal{T}$ is isomorphic to $222$ and the coset representatives are given by
\begin{align}
  T_1 &= (E,0,0,0),\label{eqt1} \\
  T_2 &= (C_{2x},\nicefrac{1}{2},\nicefrac{1}{2},0), \\
  T_3 &= (C_{2y},0,\nicefrac{1}{2},\nicefrac{1}{2}), \label{eqt3}\\
  T_4 &= (C_{2z},\nicefrac{1}{2},0,\nicefrac{1}{2}) \label{eqt4}.
\end{align}
Here, $E$ denotes the identity element and $C_{2x}$, $C_{2y}$ and $C_{2z}$ denote two-fold rotations (rotations by 180$^\circ$) about the Cartesian $x$-, $y$- and $z$-axis, respectively. In general, since $\mathcal{G}$ is an infinite group, it has infinitely many irreducible representations. However, due to the special structure of space groups they can be indexed by the combined index $(\vec{k},p)$, where $\vec{k}$ denotes a vector in reciprocal space and $p$ denotes an additional index running over all the allowed representations at $\vec{k}$. A degeneracy of a state with energy $E(\vec{k})$ can be expected when the associated irreducible representation $\Gamma_{\vec{k}}^p$ has dimension $d>1$ or when pairs of complex conjugate representations are present. Such a degeneracy is denoted as ``protected by the crystalline symmetry''. At the $R$ point within the Brillouin zone, the physically-irreducible representation is four dimensional~\cite{aroyo2006bilbao}. Additional spin-degeneracy leads to an 8-fold degenerate linear crossing. Such crossings were recently referred to as double Dirac crossings~\cite{wieder2016double}. Within the band structure in Fig.~\ref{f1d}, these crossings are highlighted by green dashed circles.

\begin{table}[b!]
\caption{{\bf Character tables and compatibility relations.}}
\subfloat[$222$ ($D_{2}$)\label{T1}]{
\begin{tabular}{lrrrr|ccc}
\hline
\hline
& $E$ & $C_{2x}$ & $C_{2y}$ & $C_{2z}$  & $\{E, C_{2x} \}$ & $\{E, C_{2y} \}$ & $\{E, C_{2z} \}$ \\
\hline
$A$ & 1 & 1 & 1 & 1 & $\Gamma_1$ & $\Gamma_1$  & $\Gamma_1$ \\ 
$B_1$ & 1 & -1 & -1 & 1 & $\Gamma_2$ & $\Gamma_2$ & $\Gamma_1$ \\
$B_2$ & 1 & -1 & 1 & -1 & $\Gamma_2$ & $\Gamma_1$ & $\Gamma_2$ \\
$B_3$ & 1 & 1 &-1 & -1 & $\Gamma_1$ & $\Gamma_2$  & $\Gamma_2$\\
\hline\hline
\end{tabular}}
\hfill
\raisebox{0.3cm}{\subfloat[$2$ ($C_2$).\label{T2}]{
\begin{tabular}{lrr}
\hline
\hline
& $E$ & $C_{2}$ \\
\hline
$\Gamma_1$ & 1 & 1 \\ 
$\Gamma_2$ & 1 & -1 \\
\hline\hline
\end{tabular}}}
\hspace{1cm}
\subfloat[$P2_12_12_1$ at $X$, $Y$ and $Z$\label{T3}]{
\begin{tabular}{lrrrr}
\hline
\hline
& $T_1$ & $T_{2}$ & $T_3$ & $T_{4}$ \\
\hline
$E$ & 2 & 0 & 0 & 0 \\
\hline\hline
\end{tabular}}
\end{table}

At the $\Gamma$ point, the group of the $\vec{k}$-vector is given by the whole space group. Each of the eigenstates is one-fold degenerate and belongs to one of the irreducible representations listed in Table~\ref{T1} (spin degeneracy is omitted for the moment). As soon as one moves slightly away from the $\Gamma$ point, for instance on the path $\overline{\Gamma X}$, the little group of the $\vec{k}$-vector only contains the elements $T_1$ and $T_2$. Hence, the bands will be classified by their transformation behavior with respect to the two-fold rotation, namely even ($\Gamma_1$) or odd ($\Gamma_2$). Every state at $X$ transforms as the 2-dimensional irreducible representation $E$ illustrated in Table~\ref{T3}. Moving towards $X$ and coming from $\overline{\Gamma X}$ bands of character $\Gamma_1$ and $\Gamma_2$ have to merge pairwise. In general, bands can only cross (accidental crossings) when they belong to different irreducible representations~\cite{herring1937}. Otherwise they will hybridize and form a spectral gap. Clearly, a similar consideration holds along $\overline{\Gamma Y}$ and $\overline{\Gamma Z}$. Taking into account all possible permutations of the four irreducible representations at $\Gamma$ as well as the possible connections to $X$, $Y$ and $Z$, it can be verified that at least one crossing can be found along one of the 3 paths. Taking into account the $C_2$ rotation symmetry, a second copy of the crossing along the path $\overline{\Gamma(-X)}$ can be also found within the Brillouin zone. Furthermore, as soon as one slightly departs from one of the 3 paths towards the interior of the Brillouin zone, none of the symmetries is kept and only the identity element $T_1$ is present. Hence, there is no reason to protect the crossing at any point that is not lying on one of the 3 paths and the crossing itself is a Dirac point. Hence, by including spin-degeneracy, this crossing is 4-fold degenerate. Moreover, it is possible to show that the $R$ point carries a spin-polarized Chern number of $C_{s}=+2$ which, by the Nielsen-Ninomiya theorem~\cite{Nielsen1,Nielsen2}, guarantees the existence of other degeneracies with a total canceling charge of $-2$. Thus, the two Dirac points must each carry a topological charge $C_{s}=-1$~\cite{ABouhonABS}.

In the present case of C$_{6}$H$_{7}$ClO$_{3}$, the topologically protected crossing can be found along the path $\overline{\Gamma X}$, as can be verified from Fig.~\ref{f1d} and Fig.~\ref{f1e}. In Fig.~\ref{f1d}, the topologically protected crossing is highlighted by a red dotted circle. As can be seen, the four interesting bands below the Fermi level have the ordering \hl{$B_3$, $A$, $B_2$, $B_1$}. Bands originating from $A$ and $B_1$ and from $B_3$ and $B_2$ are merging pairwise at $X$ for the above discussed reason. As a consequence, a crossing of bands originating from \hl{$A$ and $B_2$} can be observed (see Fig.~\ref{f1e}).

\hl{ From a generalized study of the global band topology it has been shown that more than four bands can be connected in a non trivial way. While in the case of C$_{6}$H$_{7}$ClO$_{3}$ the energy ordering of the irreducible representations at $\Gamma$ excludes an eight-band subspace nontrivially connected (realizing the global band topological class $\Gamma_I$ of }~\cite{ABouhonABS}\hl{). It is close to a topoligical transition under a band inversion between $B_1$ and $A$, as they are almost degenerate at around $-0.39$ eV} (see Fig.~\ref{f1e}). \hl{Interestingly, our data mining has found a clear candidate of an eight-band subspace nontrivially connected (realizing the global band topological class $\Gamma_{II}$ of }~\cite{ABouhonABS}\hl{) which is characterized by two pairs of topologically stable band-crossing points along $\overline{\Gamma X}$ and $\overline{\Gamma Y}$ shown in Fig.}~\ref{f3}(b). \hl{This is the first realistic band structure reported with this new type of global band topology.}

%--------------------------------
\section*{Discussion}
%--------------------------------

We presented the first 6 predicted compounds for three-dimensional organic crystals hosting isolated Dirac crossings. All 6 materials were synthesized before, so we encourage direct experimental verification of the results presented here. Within the respective band structures, we identified 8-fold degenerate Dirac nodes at the $R$ point in the Brillouin zone together with 4-fold degenerate topologically protected Dirac points along the high-symmetry path $\overline{\Gamma X}$. The crossings are well separated in energy due to the flat electronic bands of organic crystals and potentially accessible via doping or gating. In comparison to inorganic materials, the spatially sparse unit cells of organic materials characterized by van der Waals bonding between large molecules leads to an electronic structure characterized by blocks of well separated flat bands. This particular property also increases chances of finding topologically protected crossings to be isolated within the energy. We expect that more organic Dirac materials will be reported elsewhere with growth of the OMDB database. Furthermore, the slope of the crossings is usually much smaller then for similar inorganic crystals leading to potential applications as slow Dirac materials~\cite{triola2015many}. 

%--------------------------------
\section*{Methods}
%--------------------------------

%--------------------------------
% Data mining
%--------------------------------

{\bf Materials data and data mining.}
\begin{figure}[b!]
 \includegraphics[width=0.35\textwidth]{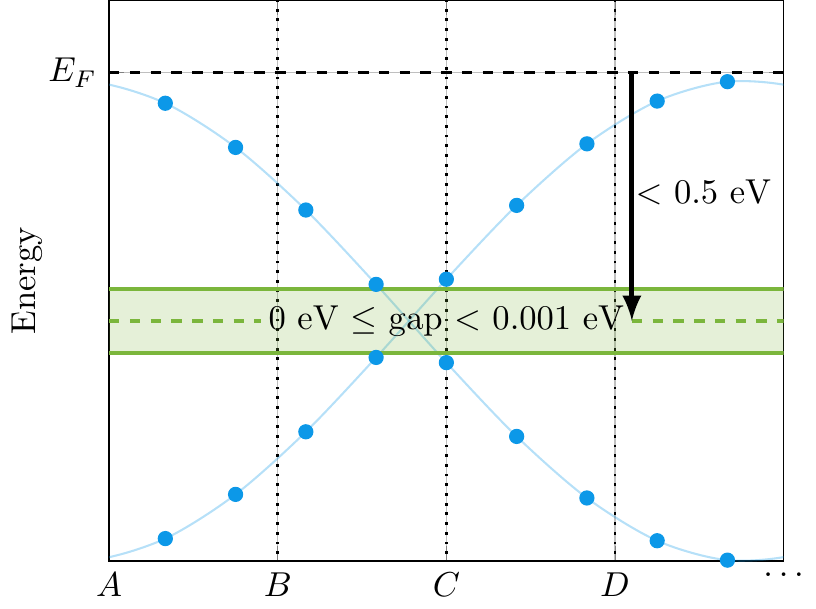}
 \caption{{\bf Illustration of the search criteria for isolated Dirac crossings near the Fermi level.} \hl{The algorithm selected all materials with zero or tiny direct energy gaps (less than 1~meV) located up to 0.5~eV above the minimum energy of the lowest conductance band or 0.5~eV below the maximum energy of the highest valence band, where no other bands can be found withing the corresponding energy range of the gap. The purpose of this gap is to introduce numerical tolerance since the band structure calculations were performed along a discrete mesh.}\label{f1}}
\end{figure} 
\hl{We analyzed 5217 Kohn-Sham band structures of the three-dimensional organic crystals stored within the Organic Materials Database (OMDB)}~\cite{borysov2016} (\url{http://omdb.diracmaterials.org}). \hl{The search algorithm for isolated Dirac crossings consisted of two major steps. On the first step, illustrated in Fig.}~\ref{f1}\hl{, the algorithm selected all materials with either zero or tiny direct energy gaps (less than 1~meV) located up to 0.5~eV above the minimum energy of the lowest conductance band or 0.5~eV below the maximum energy of the highest valence band (for metals, the reference point was the calculated Fermi energy). The purpose of this gap is to introduce numerical tolerance since the band structure calculations were performed along a discrete mesh and the algorithm could miss the crossing point if it occurs between two mesh points. The algorithm also checked that no other bands can be found within the corresponding energy range of the gap, which is a necessary (but not sufficient) criterion to find an isolated Dirac crossing. The first step allowed us to significantly reduce the search space as it found only 45 gaps (45 materials or 0.9\% of the initial dataset) near the lowest conductance band and 92 gaps (91 materials or 1.8\% of the initial dataset) near the highest valence band, making manual inspection of the search results feasible. The statistics suggest that this simple criterion performs well in filtering out irrelevant materials and can be used for the automated search for isolated Dirac crossings.} 

\hl{In the second step, we applied a pattern matching algorithm to arrange the selected materials according to their similarity to a linear crossing pattern (i.e., crossing of two straight lines). With this aim, we considered  the two nearest bands within a momentum window around a direct gap detected in the previous step. We empirically set the size of the momentum window to be 0.4 units wide as an expected characteristic scale of the Dirac crossing. Finally, we used the average Euclidean distance (the root mean square error) between the pattern and these two bands (within the momentum window) linearly scaled to the same bounding box to arrange the selected materials. The second step becomes important to prioritize search results when the number of materials is large.}

{\bf Electronic structure calculations.} Having found a subset of perspective materials, we performed refined electronic structure calculations in the framework of the density functional theory~\cite{Hohenberg1964} by applying a pseudopotential projector augmented-wave method~\cite{blochl1994projector,pseudo1,pseudo2}, as implemented in the Vienna Ab initio Simulation Package (VASP)~\cite{vasp1,vasp2,kresse1999ultrasoft} and the Quantum ESPRESSO code~\cite{qespresso}. The exchange-correlation functional was approximated by the generalized gradient approximation according to Perdew, Burke and Ernzerhof~\cite{perdew1996}. The structural information were taken from the Crystallography Open Database (COD)~\cite{merkys2016cod,gravzulis2015computing,gravzulis2012crystallography} and transformed into DFT input files by applying the Pymatgen package~\cite{Ong2013314}. 

Within VASP, the precision flag was set to ``normal'' meaning that the energy cut-off is given by the maximum of the specified maxima for the cut-off energies within the POTCAR files (for example, for carbon this value is given by 400~eV). The calculations were performed spin-polarized but without spin-orbit coupling. For the integration in $\vec{k}$-space, a $6\times6\times6$ $\Gamma$-centred mesh according to Monkhorst and Pack~\cite{monkhorst1976special} was chosen during the self-consistent cycle. A structural optimization was performed by allowing the ionic positions, the cell shape and the cell volume to change ($\textsc{isif}=3$). Optimized structures from the VASP calculations were used. Quantum ESPRESSO was applied to estimate the associated irreducible representations of the energy levels within the band structure. The cut-off energy for the wave function was chosen to be 48~Ry and the cut-off energy for the charge density and the potentials was chosen to be 316~Ry. The calculated band structures using VASP and Quantum ESPRESSO are in perfect agreement. 

\section*{Acknowledgements}
The work at Los Alamos is supported by the US Department of Energy, BES E3B7. Furthermore, the work was supported by the Swedish Research Council Grant No.~638-2013-9243, the Knut and Alice Wallenberg Foundation, and the European Research Council under the European Union’s Seventh Framework Program (FP/2207-2013)/ERC Grant Agreement No.~DM-321031, \hl{Dr. Max R\"ossler, the Walter Haefner Foundation and the ETH Zurich Foundation}. The authors acknowledge computational resources from the Max Planck Institute of Microstructure Physics in Halle (Germany) and the Swedish National Infrastructure for Computing (SNIC) at the National Supercomputer Centre at Link\"oping University.

\begin{figure*}[h!]
\subfloat[C$_{10}$H$_{10}$Br$_{2}$Cl$_{3}$NO$_{2}$]{\includegraphics[height=4.8cm]{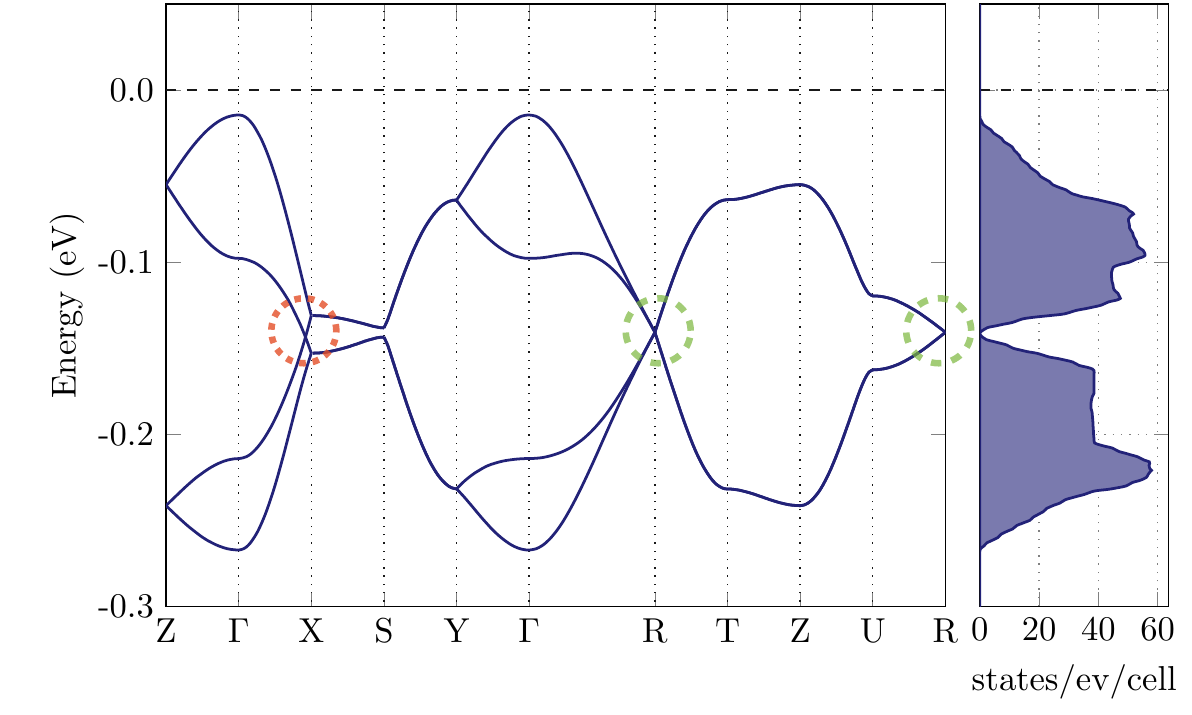}\hspace{1.cm}
\raisebox{0.4cm}{\includegraphics[height=3.5cm]{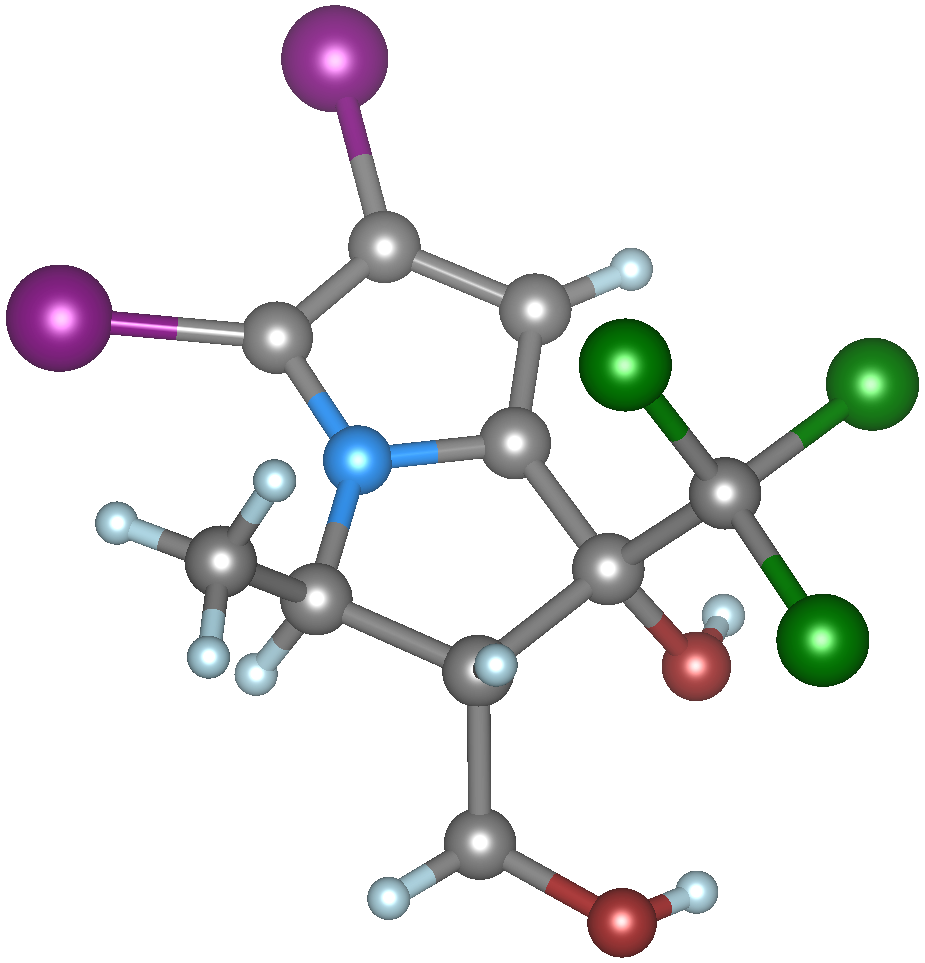}}}\\
 \subfloat[C$_{12}$H$_{13}$NO$_{2}$]{\includegraphics[height=4.8cm]{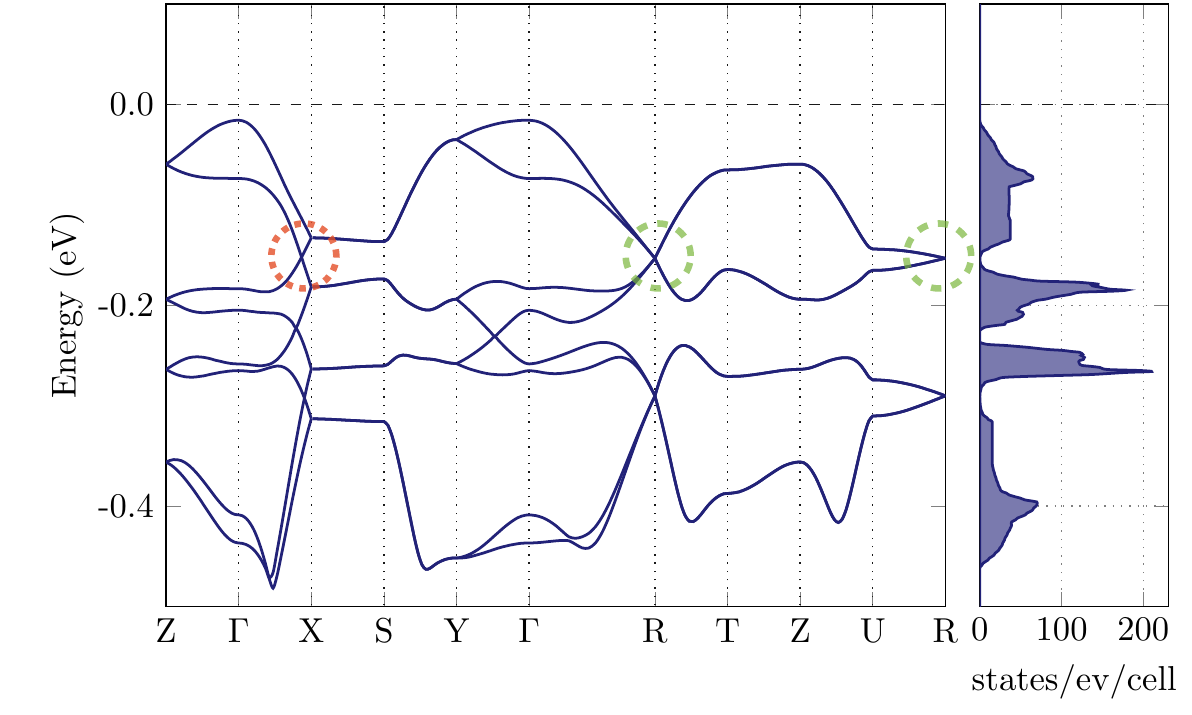}\hspace{1.5cm}
\raisebox{0.4cm}{ \includegraphics[height=3.5cm]{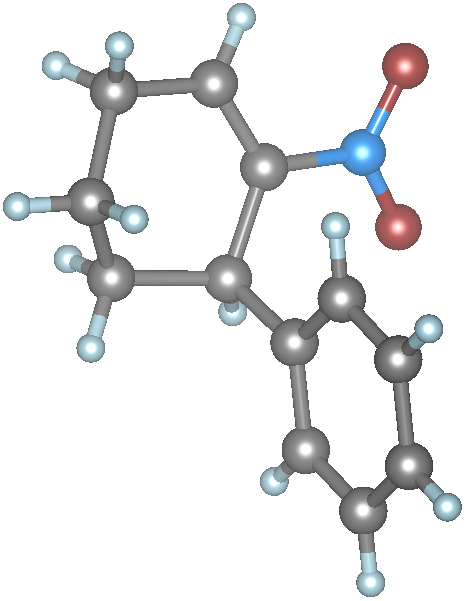}}}\\
 \subfloat[C$_{13}$H$_{12}$N$_{2}$O]{\includegraphics[height=4.8cm]{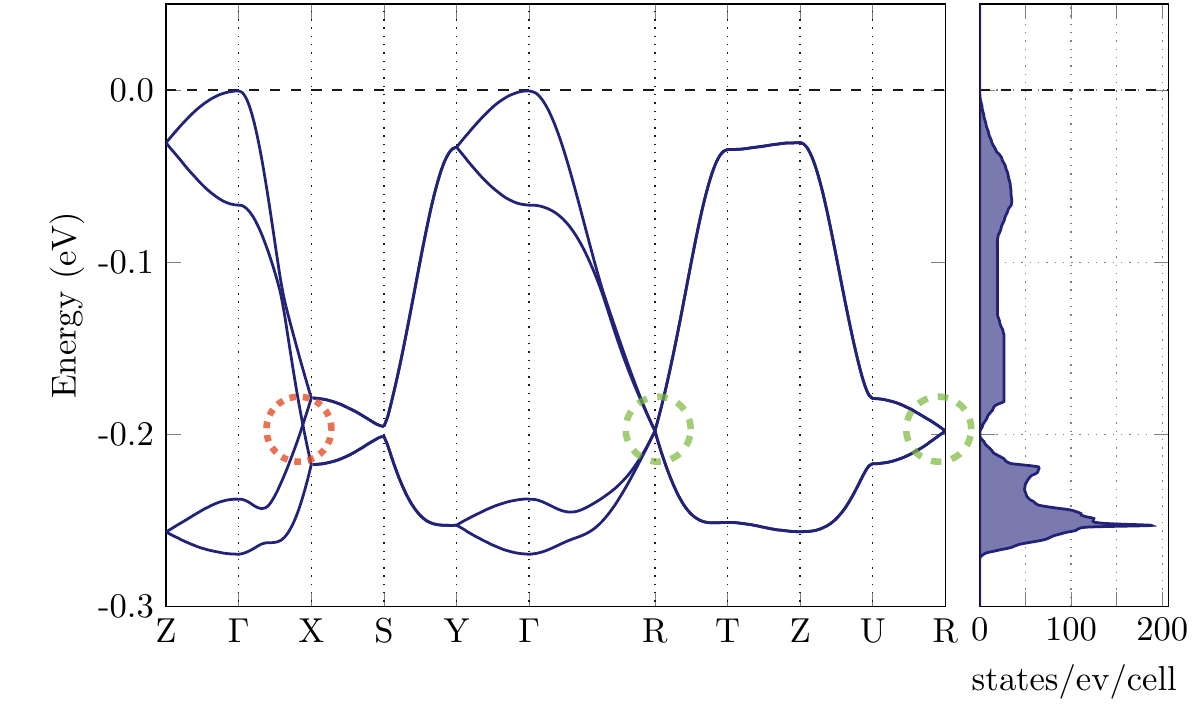}\hspace{0.4cm}
 \raisebox{0.4cm}{\includegraphics[height=3.5cm]{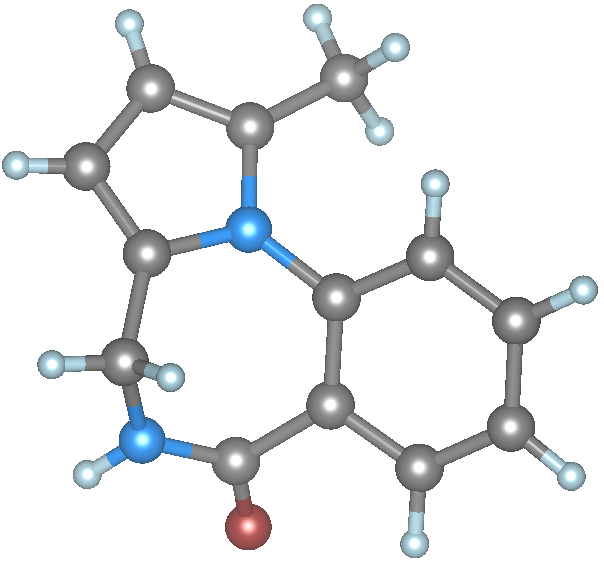}}}
\caption{\textbf{Electronic structure and molecular structure of materials with Dirac crossings in the valence band.} The colors indicate: gray: carbon, light-blue: hydrogen, blue: nitrogen, red: oxygen, purple: bromine, green: chlorine.\label{f3}}
\end{figure*}

\section*{Extended figures}
\begin{figure*}[t!]
\subfloat[C$_{9}$H$_{10}$F$_{3}$NO]{\includegraphics[height=4.8cm]{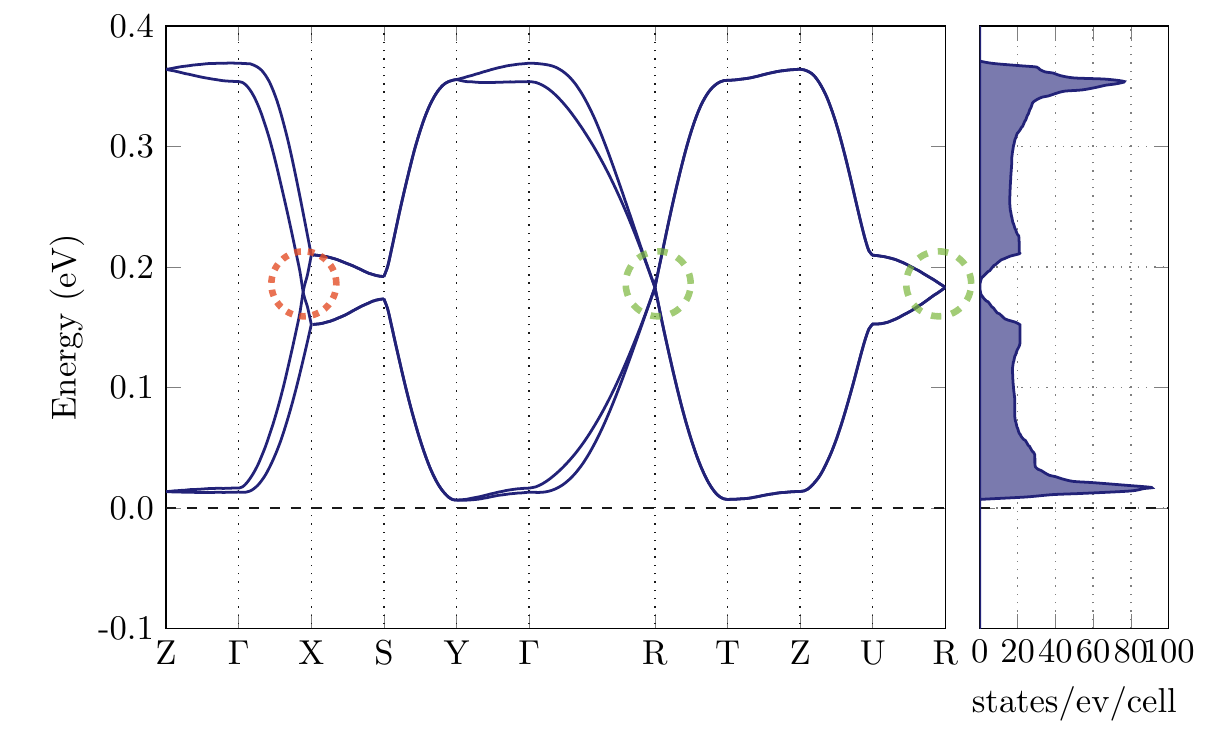}\hspace{1.5cm}
\raisebox{0.4cm}{\includegraphics[height=3.9cm]{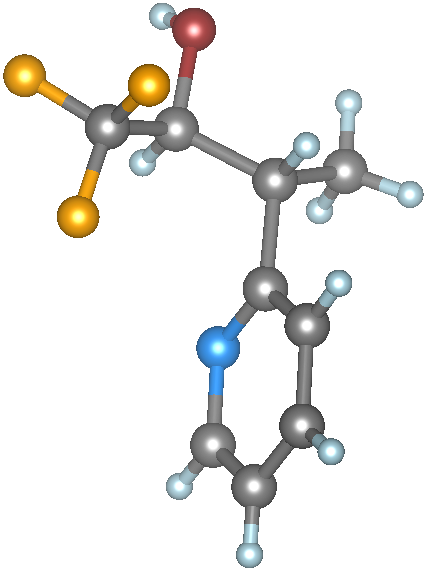}}}\\
  \subfloat[C$_{10}$H$_{12}$BrNO]{\includegraphics[height=4.8cm]{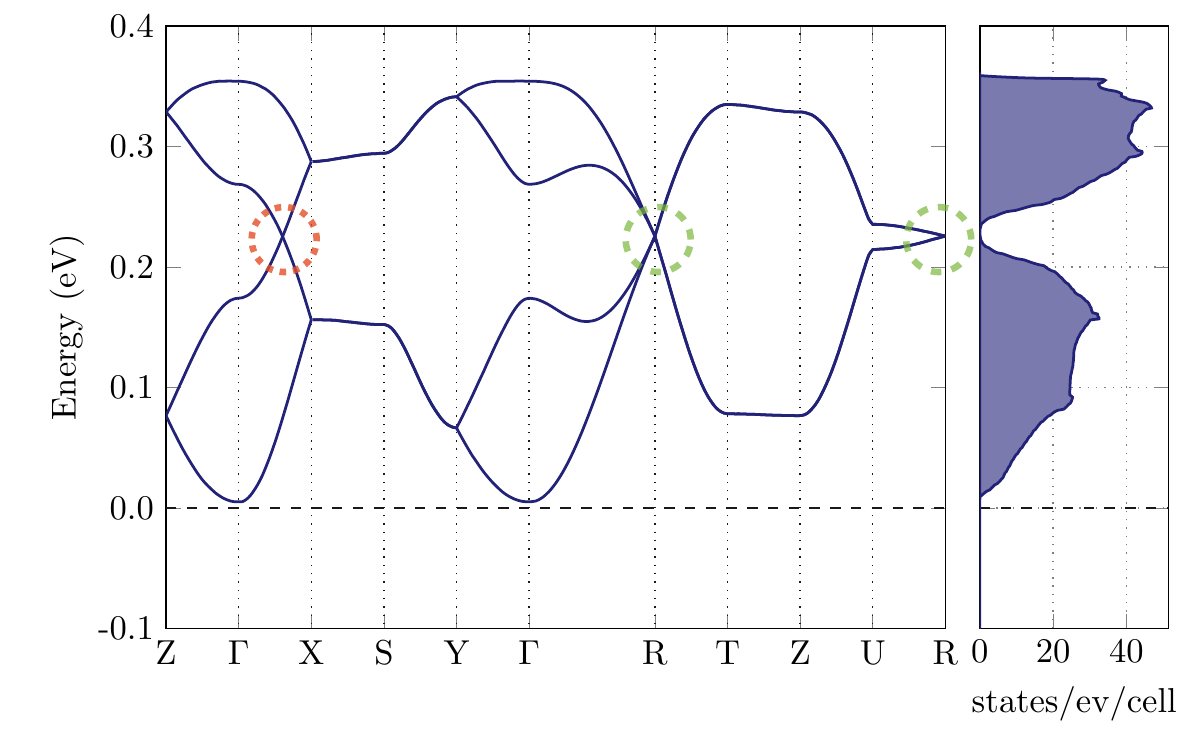}\hspace{0.5cm}
 \raisebox{0.4cm}{ \includegraphics[height=3.5cm]{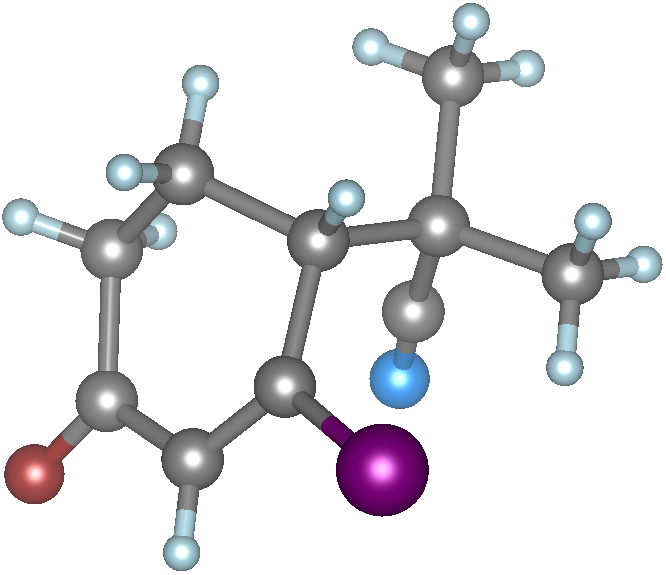}}}
\caption{\textbf{Electronic structure and molecular structure of materials with Dirac crossings in the conduction band.} The chemical elements are indicated by the colors: carbon (gray), hydrogen (light-blue), nitrogen (blue), oxygen (red), fluorine (yellow) and  bromine (purple).\label{f4}}
\end{figure*}

In the following, the band structures and molecular structures of the mined materials are presented which were not discussed in the paper so far. The materials  C$_{10}$H$_{10}$Br$_{2}$Cl$_{3}$NO$_{2}$~\cite{bae2010organocatalytic}, C$_{12}$H$_{13}$NO$_{2}$~\cite{dong2005asymmetric}, and C$_{13}$H$_{12}$N$_{2}$O~\cite{butin2010furan} contain the above mentioned Dirac crossings in the valence band close to the highest occupied electronic state and are plotted in Fig.~\ref{f3}. The distance of the crossing to the Fermi-level is in the range of 100-200 meV. Within the materials C$_{9}$H$_{10}$F$_{3}$NO~\cite{vsterk2006highly} and C$_{10}$H$_{12}$BrNO ~\cite{eloi2010cationic}, similar crossings can be found in the conduction band. The distance of the crossing to the lowest unoccupied state is about 200~meV, as can be seen in Fig.~\ref{f4}

\bibliography{references}

\end{document}